\documentclass[12pt]{article}
\usepackage{amsmath,amsfonts,amssymb,amsthm,amstext,amscd,eucal}
\usepackage[all]{xy}

\makeatletter \@addtoreset{equation}{section}

\makeatletter\renewcommand\section{\@startsection {section}{1}{\z@}%
                                   {-3.5ex \@plus -1ex \@minus -.2ex}
                                   {2.3ex \@plus.2ex}%
                                   {\normalfont\large\bfseries}}
\renewcommand\subsection{\@startsection{subsection}{2}{\z@}%
                                     {-3.25ex\@plus -1ex \@minus -.2ex}%
                                     {1.5ex \@plus .2ex}%
                                     {\normalfont\bfseries}}

\newcommand{\be}{\begin{equation}}
\newcommand{\ee}{\end{equation}}
\newcommand{\beq}{\begin{eqnarray}}
\newcommand{\eeq}{\end{eqnarray}}
\newcommand{\bea}{\begin{eqnarray}}
\newcommand{\eea}{\end{eqnarray}}
\newcommand{\fixme}[1]{\textbf{FIXME: }$\langle$\textit{#1}$\rangle$}

\def\[{\left [}
\def\]{\right ]}
\def\({\left (}
\def\){\right )}

\def\r2{\sqrt{2}}

\newcommand{\U}{\mathrm{U}}

\newcommand{\SL}{\mathrm{SL}}

\newcommand{\RR}{\mathbb{R}}

\def\sst#1{{\scriptscriptstyle #1}}

\def\1{{\sst{(1)}}}

\newcommand{\ads}[1]{${\rm AdS}_{#1}$}

\newcommand{\bbibitem}[1]{\bibitem{#1}\marginpar{#1}}

\def\Label#1{\label{#1}%
  \smash{\hbox to0pt{\raise1ex\hbox{\tiny[#1]}\hss}}}
\def\noLabels{\let\Label=\label}
\def\nobbibitem{\let\bbibitem=\bibitem}

\begin{document}
\noLabels 
\nobbibitem 

\begin{titlepage}

\begin{flushright}\vspace{-2cm}
{\small
{\tt arXiv:0906.3272} \\
IPM/P-2009-022\\
UPR-T-1209
 }\end{flushright} \vspace{5 mm}


\begin{center}

\centerline{{\Large{\bf{What is a chiral 2d CFT?}}}} \vspace{4mm}
\centerline{{\large{\bf{And what does it have to do with extremal
black holes? }}}} \vfil \vspace{6mm}

{\large{{\bf Vijay Balasubramanian\footnote{e-mail:
vijay@physics.upenn.edu}$^{,a}$,  Jan de Boer\footnote{e-mail:
J.deBoer@uva.nl}$^{,b}$,  M.M. Sheikh-Jabbari\footnote{e-mail:
jabbari@theory.ipm.ac.ir}$^{,c,d}$ \\
and Joan Sim\'on\footnote{e-mail: J.Simon@ed.ac.uk}$^{,e,f}$ }}}
\\

\vspace{5mm}

\bigskip\medskip
\begin{center} {$^a$ \it David Rittenhouse Laboratory, University
of Pennsylvania, Philadelphia, PA 19104, USA}\\
\smallskip
{$^b$ \it Instituut voor Theoretische Fysica, Valckenierstraat 65,\\
1018XE Amsterdam, The Netherlands}\\
\smallskip
{$^c$ \it School of Physics, Institute for Research in Fundamental
Sciences (IPM) ,\\ P.O.Box 19395-5531, Tehran, Iran}\\
{$^d$ \it The Abdus Salam ICTP, Strada Costiera 11, 34014, Trieste,
ITALY}\\
\smallskip
{$^e$ \it School of Mathematics and Maxwell Institute for Mathematical Sciences,\\
King's Buildings, Edinburgh EH9 3JZ, United Kingdom}\\
\smallskip
{$^f$ \it Kavli Institute for Theoretical Physics, \\
University of California, Santa Barbara CA 93106-4030, USA}\\
\end{center}
\vfil

\end{center}
\setcounter{footnote}{0}

\begin{abstract}
\noindent The near horizon limit of the extremal BTZ black hole is a
``self-dual orbifold'' of AdS$_3$.   This geometry has a null circle
on its boundary, and thus the dual field theory is a Discrete
Light Cone Quantized (DLCQ) two dimensional CFT.   The same geometry can be
compactified to two dimensions giving AdS$_2$ with a constant
electric field.   The kinematics of the DLCQ show that in a
consistent quantum theory of gravity in these backgrounds there can be no
dynamics in  AdS$_2$, which is consistent with older ideas about
instabilities in this space.  We show how the necessary boundary
conditions eliminating AdS$_2$ fluctuations can be implemented,
leaving one copy of a Virasoro algebra as the asymptotic symmetry
group. Our considerations clarify some aspects of the chiral CFTs
appearing in proposed dual descriptions of the near-horizon degrees
of freedom of extremal black holes.
\end{abstract}
\vspace{0.5in}

\end{titlepage}
\renewcommand{\baselinestretch}{1.05}  
\tableofcontents


\section{Introduction}

In the vicinity of their horizons, extremal black holes in many
dimensions, both in flat and anti-de Sitter spaces, contain an
\ads{2} component with a constant electric field\footnote{This
statement is an actual theorem in four and five dimensions, under
certain isometry assumptions and for extremal black holes with finite
area horizons \cite{harvey}.}.   Proposed dualities between \ads{2}
space and a conformal quantum mechanics
\cite{MAGOO,Spradlin:1999bn,confqm,Sen3to2, tak} or a chiral 1+1 dimensional
conformal field theory (CFT) \cite{Sen3to2, AdS2QGr} have been used
to explain the statistical degeneracy of extremal black holes.   In
\cite{AdS2QGr,Vijay-Asad-Joan} it was shown the \ads{2} geometry
with a constant electric field can be understood as the
compactification of an orbifold of \ads{3} with a null boundary.
Systematically applying the rules of the AdS/CFT correspondence then
suggests that the dual theory on the 1+1 dimensional boundary is a
Discrete Light Cone Quantized CFT
\cite{AdS2QGr,Vijay-Asad-Joan,Semenoff,tak}.   Because of the highly
boosted kinematics of a DLCQ theory, only one chiral sector of the
$2d$ CFT survives.  Such chiral theories thus seem to appear
universally in the dual descriptions of extremal black holes.

In this paper, we develop aspects of this DLCQ - extremal black hole
correspondence.  The essential features can be understood by
considering the extremal BTZ geometry, which itself appears in the
near-horizon geometry of many asymptotically flat or AdS black
holes.   It is well known that the BTZ black holes are dual to
thermal ensembles in a 1+1 dimensional CFT.   Thermal ensembles in a
single chiral sector of this CFT are dual to the extremal black
holes and explain their statistical degeneracy.  Taking a limit
which focuses on the vicinity of the BTZ horizon gives a locally
\ads{3} geometry that is a circle fibration over an \ads{2} base.
{}From the three dimensional perspective this is precisely the
self-dual orbifold of \cite{Vijay-Asad-Joan,Coussaert:1994tu}.
Dimensionally reducing over the circle fibre gives an \ads{2}
geometry with an electric flux -- precisely the spacetime appearing
in \cite{AdS2QGr}. As we will show in sections \ref{BTZ-dual-CFT}
and \ref{DLCQ-section}, the same focusing limit applied to the CFT
dual to BTZ effectively applies a DLCQ procedure that isolates the
chiral sector carrying the extremal black hole entropy.  Thus, one
chiral set of Virasoro generators of the CFT is frozen in this
limit, in the sense that there are no physical states charged under
them. It turns out that the same chiral sector also contains the
$\SL(2,\RR)$ isometries of the \ads{2} geometry, while the surviving
$\SL(2,\RR)$ in the limiting chiral CFT appears as an enhancement of
the $\U(1)$ symmetry of the circle fibration. Specifically, we show
that there exists a consistent set of boundary conditions on the
fluctuations of the near horizon extremal BTZ metric, as in the
Brown-Henneaux analysis \cite{Brown-Henneaux}, that enhances the
$\U(1)$ isometry to an asymptotic chiral Virasoro algebra. This is
consistent with recent proposals that the description of extremal
black holes in terms of an \ads{2} throat requires asymptotic
boundary conditions eliminating \ads{2} excitations and enhancing a
$\U(1)$ appearing in the geometry to a Virasoro symmetry
\cite{Kerr/CFT,Hartman:2008pb}.

Usually in the AdS/CFT duality, the isometries of spacetime are
realized in the dual as global symmetries which then organize the
representations of physical states.  The surprise here is that the
$\SL(2,\RR)$ symmetry inherited in the CFT from the spacetime
isometries acts trivially on the space of physical states. This  has
two implications.  First, the chiral duals to the near-horizon geometry
of extremal black holes are incapable of describing non-extremal
excitations. Second, even after the addition of an electric field to
\ads{2}, $2d$ quantum gravity with this asymptotics has no dynamics.
This is consistent with the idea that finite energy excitations in
\ads{2} destroy its asymptotic structure \cite{AdS2-fragment}. These
two points are related to the fact that non-extremal black holes do
not have \ads{2} throats.  Similarly, in the classic setting of the
D1-D5-string, extremal black holes arise from chiral excitations,
and non-extremality requires excitations of both left and right
movers.

The self-dual orbifold and \ads{2} with a flux also appear in the
near horizon limit of the extremal Kerr black hole in four
dimensions \cite{Kerr/CFT} suggesting the appearance of a chiral CFT
dual.  However, in this setting (as in
\cite{Balasubramanian:2007bs,Fareghbal:2008ar,Fareghbal:2008eh}) the
near-horizon AdS geometries appear in a ``warped'' way, with their
metric multiplied by a function of another angular direction in the
overall spacetime.  We suggest that reduction over this additional
direction can give rise to an effective three dimensional gravity
with a negative cosmological constant with the self-dual orbifold as
a solution. The dual description of this space as a chiral $2d$ CFT
then explains the statistical degeneracy of Kerr.

\emph{Note added:} In the last stage of preparation of this article
two papers appeared on the arXiv \cite{No-dynamics-NHEK} arguing
that there is no dynamics in the  chiral $2d$ CFT proposed to be
dual to the near horizon extremal Kerr geometry, in agreement with
our results.

\section{Near horizon extremal BTZ is dual to  DLCQ of a $2d$ CFT}
\Label{BTZ-dual-CFT}%

BTZ black holes are three dimensional, asymptotically AdS$_3$ spacetimes with metric  \cite{BTZ-original}%
\be\label{BTZ-metric}%
ds^2 = -\frac{(r^2-r_+^2)(r^2-r_-^2)}{r^2 \ell^2} dt^2
 +  \frac{\ell^2 r^2}{(r^2-r_+^2)(r^2-r_-^2)} dr^2
 + r^2 (d\phi - \frac{r_+r_-}{\ell r^2} dt)^2 \,.
\ee%
They have ADM angular momentum and mass%
\be\label{BTZ-MJ}%
 \frac{J}{2}
=\frac{r_+ r_-}{\ell^2}, \qquad M=\frac{r_+^2 + r_-^2}{\ell^2}%
\ee%
given in terms of two parameters: the inner and outer horizons
$r_\pm$. These are locally AdS$_3$ spacetimes, differing from global
AdS$_3$ by a quotient under a discrete identification. This is the
origin of the periodicity in $\phi$ in \eqref{BTZ-metric}, i.e.
$\phi \sim \phi + 2\pi$. Regularity of the metric requires $|J| \leq
M $. The BTZ black holes also appear as components in the
near-horizon geometry of black holes in many dimensions with both
vanishing and negative cosmological constants (e.g. see
\cite{AdS3-reviews}). The {\it extremal} BTZ black holes ($M=J$)
have coincident inner and outer
horizons%
 \be%
  M = J ~~~~\Longrightarrow~~~~ r_+=r_-  \equiv r_h \, .
\ee%
 Globally,  the generator of the discrete quotient of \ads{3}
giving rise to the extremal black hole lies in a different conjugacy
class from the generator giving rise to the non-extremal black hole
\cite{Banados:1992gq}.

According to the AdS/CFT correspondence, quantum gravity in AdS$_3$
is dual to a $2d$ conformal field theory (CFT) with
equal left and right central charges $c$ \cite{Brown-Henneaux}%
\be\label{B.H-central-charge}%
c=\frac{3\ell}{2G_3}\,,%
\ee%
where $G_3$ is Newton's constant in three dimensions.  The BTZ black
holes are thermal states in this CFT having left and right-moving
temperatures
\be\label{BTZ-temperature}%
T_{R} = \frac{1}{4\pi} \frac{r_+ - r_-}{\ell}\ ,
\qquad %
T_{L} = \frac{1}{4\pi} \frac{r_+ + r_-}{\ell}\ ,
\ee%
with energy and angular momentum:%
\be\label{L0-barL0}%
L_0-\frac{c}{24}= M-J,\qquad {\bar
L}_0-\frac{c}{24}=M+ J.%
\ee%
(In our conventions $M$ and $J$ are both dimensionless; their
natural units are given by the AdS$_3$ radius $\ell$.)     In the
extremal ($M=J$) black hole the right-movers are  in the ground
state\footnote{In theories with supersymmetry these are indeed the
obvious ground states in the RR sector. This condition can even
correspond to ground states in the NS sector, because in many
examples the quantum numbers $L_0$ and $\bar{L}_0$ are not exactly
identical to the standard CFT quantum numbers but can e.g. receive
contributions from gauge fields which make them spectral flow
invariant, in which case this condition  really implies that the
states have to be chiral primary. Although we have no proof that
$L_0=c/24$ always implies that the states have to be ground states
of some sort, we will continue to refer to these states as ground
states and hope that this will not cause any confusion.}
\be%
L_0 = {c \over 24} ~;~~~~ T_R = 0%
\ee%
while the left moving temperature $T_L = {1 \over 2\pi} {r_h \over
\ell}$ and $\bar{L}_0$ are arbitrary.  The extremal BTZ entropy
(and that of higher dimensional black holes of which it is the
near horizon limit) is accounted for by the statistical degeneracy
of such a chiral CFT sector with $\bar{L}_0 - c/24 = 2M$, at least
when $\bar{L}_0 \gg c/24$.

%


We will see that the chiral sector that is responsible for the
extremal entropy can be isolated by taking a near-horizon limit of
the extremal BTZ black hole \cite{AdS2QGr}.   It is convenient to do
this
in another set of coordinates \cite{Carlip}%
\be\label{uvrho}%
\hat u=t/\ell-\phi\ ,\qquad \hat v=t/\ell+\phi,\qquad
r^2-r_+^2=\ell^2
e^{2\rho}\,, %
\ee%
in which the metric takes the form  \footnote{For later use
note that a generic BTZ metric in the $\hat u, \hat v, \rho$ coordinate
system takes the form \cite{Carlip}%
\be\label{BTZ-uv-ccordinates}%
ds^2=\ell^2\left[L^+ d\hat u^2+ L^- d\hat v^2+ d\rho^2-(e^{2\rho}+L^+L^-e^{-2\rho}) d\hat ud\hat v\right]\ .%
\ee%
where $ L^\pm=\frac{1}{4\ell^2}(r_+\pm r_-)^2.$ Recalling
\eqref{BTZ-temperature}, $L^+=(2\pi T_L)^2,\ L^-=(2\pi T_R)^2$.
}%
\be%
ds^2=r^2_+ \,  d\hat u^2+\ell^2 \,  d\rho^2-\ell^2 e^{2\rho} \, d\hat u \,  d\hat v\ .%
\Label{Ext-BTZ}
\ee%
The variables  $\hat u,\ \hat v$ have a periodicity
\be
\{\hat{u},\hat{v}\} \sim \{\hat{u} - 2\pi, \hat{v} + 2\pi \} \, .
\Label{uhatvhatperiod}
\ee
On the cylindrical boundary of \ads{3} ($\rho \to \infty$),
$d\hat{u}$ and $d\hat{v}$ become null directions.   Thus the two
chiral Virasoro algebras of the dual conformal field theory are
associated to asymptotic reparameterizations $\hat{u} \to
f(\hat{u})$ and $\hat{v} \to g(\hat{v})$.


Since the horizon is located at $\rho\to -\infty$, we take the near horizon limit%
\be%
\rho=\rho_0+ r,\quad u=\hat u \, {r_+ \over \ell},\quad v=
{e^{2\rho_0} \ell \over r_+ } \, \hat v , \quad \{ u,v \} \sim \{u -
2\pi {r_+ \over \ell}, v + 2\pi {\ell \over r_+} e^{2 \rho_0} \}
\quad (\rho_0\to-\infty)%
\Label{NHlimit-BTZ}
\ee%
while keeping $r,\ u,\ v$ and $r_+$ fixed.\footnote{Despite the
resemblance of the  limit \eqref{NHlimit-BTZ} and the coordinate
changes one makes in taking the Penrose limit, \eqref{NHlimit-BTZ}
is not a Penrose limit, as the geometry we obtain after the limit is
not a plane-wave.} (See  \cite{AdS2QGr} for the first discussion of
this limit.)
 The resulting metric, which describes the geometry in the vicinity of the extremal horizon,
\be%
 ds^2 = \ell^2(  du^2 + dr^2 - e^{2r} \, du \, dv )
\Label{NHlimit-BTZmet1}%
 \ee %
is identical in form to (\ref{Ext-BTZ})
but there is a crucial difference.  In the $\rho_0 \to -\infty$
limit, the identification (\ref{uhatvhatperiod}) becomes \be
 \{u, v \} \sim \{u - 2\pi {r_+ \over \ell}, v \} \, .
 \ee%
Thus, the boundary of (\ref{NHlimit-BTZmet1}) ($r \to \infty$) is a
``null cylinder'' -- it has a metric conformal to $du \, dv$, the
standard lightcone metric on a cylinder, but has a compact {\it
null} direction ($u$).   The periodicity of $u$ encodes the
temperature of the left-moving thermal state that gave the original
extremal BTZ black hole its statistical degeneracy.

Rewriting the radial coordinate as $y = e^{2r}$ gives
\be
ds^2 = \frac{\ell^2}{4}\left( -y^2 \,  d v^2
 +  \frac{dy^2}{ y^2}\right)+ \ell^2 \left(du - \frac{1}{2 } \, y \, dv\right)^2  .%
\Label{NH-BTZ-Sl2xU1}%
\ee%
This is an $S^1$ fibration over AdS$_2$ which arises as a discrete
identification of AdS$_3$. The generator of this discrete group
sits inside the
$\SL(2,\mathbb{R})_L$ subgroup of the initial
$\SL(2,\mathbb{R})_L\times \SL(2,\mathbb{R})_R$ isometry group of
AdS$_3$ \cite{AdS2QGr,Vijay-Asad-Joan}. To be precise, the
parametrization of $\SL(2,\mathbb R)$ (i.e. \ads{3}) that is
relevant for the metric (\ref{NH-BTZ-Sl2xU1}) is
 \be \label{coords2} {\cal G} = \left(
\begin{array}{cc} 1 & 0 \\ {v \over 2} & 1
\end{array} \right)
\left( \begin{array}{cc} \sqrt{y} & \sqrt{y} \\
\frac{-1}{2\sqrt{y}} & \frac{1}{2\sqrt{ y}}
\end{array} \right)
\left( \begin{array}{cc} e^{u} & 0 \\ 0 & e^{-u}  \end{array}
\right) \, , \ee in terms of which the metric (\ref{NH-BTZ-Sl2xU1})
is%
\be%
 ds^2 =  {\ell^2 \over 2} {\rm tr}({{\cal G}^{-1}} \, d{\cal
G} )^2 %
\ee%
Under $u \rightarrow u - 2\pi r_+ / \ell$, ${\cal G}$ is identified
by the right action of %
\be%
\left(
\begin{array}{cc} e^{-2 \pi r_+/\ell} & 0 \\ 0 &
e^{2 \pi r_+/\ell}
\end{array} \right) \, .
\ee%
The isometry group is $\SL(2,\mathbb{R})_R\times \U(1)_L$, the first
factor corresponding to the isometries of the AdS$_2$
base.\footnote{Strictly speaking, these $\SL(2,\RR)$ transformations
include $\U(1)$ gauge transformations compensating the
transformation of the gauge field on AdS$_2$.} On the boundary of
the spacetime these isometries act to reparameterize the non-compact
coordinate $v$.   In fact, this geometry is precisely the {\it
self-dual orbifold} of Coussaert and Henneaux
\cite{Coussaert:1994tu}. The present coordinate system covers only
part of the global spacetime described in \cite{Vijay-Asad-Joan,
Coussaert:1994tu}.

Since  \eqref{NHlimit-BTZmet1} is asymptotically locally \ads{3}, we
expect the dual field theory to still be a two dimensional conformal
field theory, but defined on a boundary null cylinder.  To
understand what that means, we can follow \cite{Vijay-Asad-Joan} and
regulate the CFT by cutting off the self-dual orbifold at a fixed,
large radius. Following the usual AdS/CFT reasoning, this implements
a UV cutoff in the field theory.  We will remove the cutoff by
sending $r \to \infty$.  At any fixed $r$, the metric
\eqref{NHlimit-BTZmet1} is conformal to \be ds^2 = du^2 - e^{2r} \,
du \, dv \Label{NHmetsimp} \ee Now consider a standard cylinder with
its usual Cartesian metric
$
ds^2 = -dt^2_0 + d\phi_0^2
$
and $ \{ \phi_0 , t_0 \} \sim \{\phi_0 - \beta, t_0 \}$. We will use
coordinates%
\be%
u_1 = t_0 - \phi_0 ~~~;~~~ t_1 = 2 t_0
~~~\Longrightarrow~~~ ds^2 = du_1^2 - du_1 \, dt_1 ~~~;~~~ \{
u_1,t_1 \} \sim \{ u_1 + \beta, t_1\} \, . \Label{specmet1} %
\ee%
 We now boost the cylinder with a rapidity $2\gamma$ ($\tilde{u}_1=
e^{2\gamma} u_1$) and then reparameterize  the boosted cylinder so
that the identification is still occurring at fixed $t_1$.  The
metric then becomes \be ds^2 = e^{-4\gamma} \left(d\tilde{u}_1^{2} -
e^{2\gamma} d\tilde{u}_1 \, dt_1 \right) ~~~;~~~ \{ \tilde{u}_1, t_1
\} \sim \{\tilde{u}_1 + \beta e^{2\gamma},t_1 \} \Label{specmet2}
\ee Rescaling the coordinates as $\tilde{u}_1 \to e^{-2\gamma}
\tilde{u}_1$ and $t_1 \to e^{-2\gamma} t_1$ gives the metric \be
ds^2 =d\tilde{u}_1^{2} - e^{2\gamma} d\tilde{u}_1 \, dt_1 ~~~;~~~ \{
\tilde{u}_1, t_1 \} \sim \{\tilde{u}_1 + \beta ,t_1 \} \, .
\Label{specmet3} \ee Thus,  metrics on fixed $r$ surfaces of the
near-horizon BTZ metric (\ref{NHmetsimp}) are conformal to a boosted
cylinder.   As $r \to \infty$ the boost becomes infinite, precisely
realizing the procedure defined by Seiberg \cite{Seiberg} for
realizing the Discrete Light Cone Quantization (DLCQ) of a field
theory. In Sec.~\ref{DLCQ-section} we will show that following the
usual kinematics of DLCQ, only one chiral sector of the CFT dual to
\ads{3} will survive at finite energies.

We can also see the latter  by directly examining the near-horizon
limit (\ref{NHlimit-BTZ}).  Acting in the CFT dual to \ads{3}, the
near horizon limit of the extremal BTZ black hole focuses in on
energies so low that they lie below the black hole mass gap, thus
eliminating all non-extremal dynamics \cite{AdS2QGr} (also see \cite{tak}).
This will
isolate one chiral sector (the left-movers), since non-extremal,
finite energy excitations necessarily involve excitations of the
right-movers also. Explicitly, the infinite rescaling in the
coordinate $\hat v$ relates translations as \be
\partial_v \sim
e^{-2\rho_0}\,\partial_{\hat v} \, .
\Label{scaledham}
\ee%
Thus, recalling the $\partial_{\hat{v}}$ is the right-moving
Hamiltonian in the CFT dual to \ads{3}, any finite-energy
right-moving excitation, i.e. any excitation $|s\rangle$ with
$\partial_{\hat{v}} |s\rangle= (L_0-c/24) |s\rangle \neq 0$, will be
infinitely blue shifted in the Hamiltonian $\partial_v$ that is well
defined in the $\rho_0 \to - \infty$ limit.  In other words, we
should only be keeping the states satisfying%
\be%
\partial_{\hat{v}} |s\rangle = (L_0 - c/24) |s\rangle = 0
\label{vann}
\ee%
which are the ground states in the right-moving sector.

We can also directly follow how the near-horizon limit
(\ref{NHlimit-BTZ}) acts on the left and right moving Virasoro
generators of the CFT dual to \ads{3}.  These generators are
\be%
L_n-\frac{c}{24}\delta_{n,0}=e^{in\hat v}\frac{\partial}{\partial
\hat v},\qquad {\bar
L}_n-\frac{\bar c}{24}\delta_{n,0}=e^{in\hat u}\frac{\partial}{\partial \hat u}\ .%
\Label{ads3virasoro}
\ee%
As $\rho_0 \to - \infty$ in the near-horizon limit
\eqref{NHlimit-BTZ}, it is evident that ${\bar L_n}$  are
essentially unchanged while the $L_n$ annihilate all the
\emph{finite energy} states because of the condition
(\ref{vann}).

\section{DLCQ of a $2d$ CFT is a chiral CFT}
\Label{DLCQ-section}%

In the previous section we reviewed how the near-horizon geometry
of extremal BTZ is dual to the DLCQ of a $2d$ CFT. We now examine
how such theories are quantized.  Consider a $2d$ CFT on a
cylinder
\be%
ds^2 = -dt^2 + d\phi^2 = -du' \, dv'   ~~~;~~~ u' = t - \phi,  \  v' = t + \phi
\label{cylinder}
\ee%
where $\phi$ is a circle with radius $R$.  Here
\be
\{\phi,t\} \sim \{\phi + 2\pi R, t \} ~~~~;~~~~ \{u' , v'\} \sim
\{u' - 2\pi R, v' + 2\pi R \} \label{standcyl}
\ee
Let $P^{u'}$ and $P^{v'}$ denote momentum operators in the $v'$
and $u'$ directions respectively.   Their eigenvalues
 \be%
 P^{v'} =
\left(h+n-\frac{c}{24}\right)\frac{1}{R},\qquad P^{u'} =
\left(h-\frac{c}{24}\right)\frac{1}{R}\,, \qquad  n\in \mathbb Z %
\ee%
are given in terms of the quantized momentum $n$ along the $S^1$,
the $2d$ central charge $c$ and an arbitrary value of $h$ with $h
\geq 0$ and $h + n \geq 0$.  These are related to the eigenvalues
of the standard operators $L_0,\bar{L_0}$ used in radial
quantization on the plane by $\bar L_0=h+n$ and ${L}_0 = h$. We
will assume that the $2d$ CFT is non-singular, and therefore that
the spectrum is discrete.

Following Seiberg \cite{Seiberg}, consider a boost with rapidity $\gamma$
\be\label{boost}%
u'  \to e^\gamma u', \qquad v' \to e^{-\gamma} v' \, .
\ee%
The boost leaves metric (\ref{cylinder}) invariant.\footnote{In
our previous analysis the cylinder in coordinates (\ref{specmet1})
was boosted but also reparameterized -- this is why the metric
transformed to (\ref{specmet2}).}  However the identifications are
now \be \{ u' , v' \} \sim \{u' -2\pi R e^{\gamma} , v' + 2\pi R
e^{-\gamma} \}       \, . \ee We want to match the boundary
structure appearing in the boundary of the near horizon geometry
with the DLCQ of the starting boundary cylinder. To do so,
consider the limit  $\gamma \to \infty$ with $R\,e^\gamma$ fixed.
This describes a null cylinder geometry with metric $ds^2 = - du'
\, dv'$ and $u'$  a compact null direction.  The same infinite
boost was presented in different coordinates in (\ref{specmet1})
-- (\ref{specmet3}). However, notice that since $v' \to
e^{-\gamma} v' = e^{-\gamma} (t + \phi)$ and $0 \leq \phi \leq
2\pi\,R$, as $\gamma \to \infty$ any finite changes in $v'$ come
from changes in $t$.  Thus, in the limit, $dv' \propto dt$ and
$ds^2 = - du' \, dv' \approx - e^{-\gamma} du' \, dt$ which is
conformal to the dominant piece of the metric in (\ref{specmet3}).

More explicitly, the periodicities of the boundary coordinates under
the limit  $\gamma \to \infty$ with $R_-\equiv R\,e^\gamma$ fixed
are
\be%
\left(\begin{array}{cc} \phi \\ t \end{array}\right)\sim
\left(\begin{array}{cc} \phi \\
t \end{array}\right)+\left(\begin{array}{cc} 2\pi R \\ 0
\end{array}\right) \ -{\mathrm{infinite\ boost}}
\rightarrow \ \left(\begin{array}{cc} u' \\ v'
\end{array}\right)\sim \left(\begin{array}{cc} u' \\ v'
\end{array}\right)+ \left(\begin{array}{cc} 2\pi  R_-  \\
2\pi R_- e^{-2\gamma}
\end{array}\right)
\Label{tpmy-periodicity}
\ee%
We can now identify $\{u',v'\}$ with the lightcone boundary coordinates
of \ads{3} in \eqref{NHlimit-BTZ} via $u' = u (\ell / r_+) R_-$ and
$v' = v (r_+ / \ell) R_-$.  Then, comparing (\ref{NHlimit-BTZ}) and
(\ref{tpmy-periodicity}), it is evident that the action of the near
horizon limit on $u,v$ precisely reproduces the identifications
induced by the infinite boost in DLCQ.   Thus, from this perspective
also, the dual to the near-horizon geometry of the extremal BTZ
black hole should be the DLCQ of the 1+1 dimensional CFT dual to
\ads{3}.

Because of the kinematics of the DLCQ boosts,
\be\label{Ppm-limit}%
 P^{v'} =
\left(h+n-\frac{c}{24}\right)\frac{e^{-\gamma}}{ R},\qquad P^{u'} =
\left(h-\frac{c}{24}\right)\frac{e^{\gamma}}{ R}\,. %
\ee%
Keeping $P^{u'}$ (momentum along  $v'$) finite in the  $\gamma \to \infty$ limit requires
$h=c/24$. This leads to
\be%
P^{v'} =n\cdot \frac{e^{-\gamma}}{ R}=\frac{n}{R_-}\ .%
\ee%
Thus
the DLCQ limit \eqref{Ppm-limit} freezes the right moving sector.
Equivalently, it generates an infinite energy gap in this sector,
while the gap in the left-moving sector (whose energy is measured
by $P^{v'}$) is kept finite.  All physical finite energy states in
this limit only carry momentum along the compact null direction
$u'$. Therefore, the DLCQ  $\gamma \to \infty$ limit defines a
Hilbert space ${\cal H}$
\be%
 {\cal H} = \{  |{\rm anything}\rangle_L\otimes |c/24\rangle_R\}\,.%
\ee%

It is worth noting that the extremal D1-D5-p black hole (whose
near horizon limit is the BTZ black hole) is precisely dual to
states of this form with the right movers in the RR ground state,
and the left movers in a highly excited state the  statistical
degeneracy of which explains the black hole entropy
\cite{Strominger-Vafa}.

Since the spectrum of the DLCQ theory is chiral we might wonder what
remains in this limit of the Virasoro algebra of the CFT we started
with.  Denoting the right moving Virasoro generators by $L_m$,  all
states $L_m\,|\,c/24\rangle$ ($m<0$) have infinite energy in the
DLCQ limit, since their action always changes the right-moving
energy.  Explicitly, consider the generators \be
L_q \sim e^{iq v'} \frac{\partial}{\partial v'}\ ,\qquad {\bar L}_p \sim e^{ip u'} \frac{\partial}{\partial u'} \, , %
\Label{LqLp-DLCQ}%
\ee%
with $L_0-c/24$, ${\bar L}_0-c/24$ being  generators of
translations along $v'$ and  $u'$ respectively.
After the boost \eqref{boost} the quantization conditions  for
$p,q$ become:
\be%
q = \frac{k }{ R e^{-\gamma}}=\frac{k}{R_-} e^{2\gamma},\qquad
p= {m}\cdot \frac{1}{R e^{\gamma}}=\frac{m}{R_-},\qquad
k,m\in\mathbb Z.%
\ee %
Thus, there is a single copy of the Virasoro algebra, generated by
${\bar L}_p$, which survives the limit. This is acting on the left
movers, as expected from the spectrum defining the Hilbert space of
the theory. Notice the generators of this algebra are acting on the
compact direction of the DLCQ null cylinder.

\noindent {\bf Summary: } The DLCQ of a non-singular $2d$ CFT
freezes the right moving sector to its ground states
$|\,c/24\rangle$ while keeping the full left moving sector. Hence,
the DLCQ limit gives a chiral $2d$ CFT with the same central
charge as the original one. Applied to the BTZ black hole
(Sec.~\ref{BTZ-dual-CFT}), we learn that the near-horizon geometry
of extremal BTZ is dual to one chiral sector of the $2d$ CFT with
central charge $c=3\ell/2G_3$ that is dual to \ads{3} gravity. The
surviving chiral sector is in the state in which it was placed to
realize the dual to an extremal black hole, namely a thermal state
at a temperature $T_L=T_{DLCQ}=R_-/(2\pi)$, corresponding to the
left-moving thermal state $|\,c/24\rangle  \otimes |\,T=R_-/2\pi
\rangle$ in the Hilbert space of the CFT dual to \ads{3}.

\section{Asymptotic symmetries and the chiral Virasoro algebra}

In the AdS/CFT correspondence, the isometries of  spacetime manifest
themselves as global symmetries of the dual field theory, and
physical states are organized in representations of the isometry
group.  For this reason, various authors \cite{Vijay-Asad-Joan,
Hartman, Finn-Larsen} have considered how the physical states of
fields in the near-horizon BTZ geometry (\ref{NHlimit-BTZmet1}) or
(\ref{NH-BTZ-Sl2xU1}) transform under the $\SL(2,\RR) \times \U(1)$
isometry group.    Now recall that the DLCQ analysis of the dual
field theory in the previous section showed that the physical states
of this theory must live in a chiral CFT.   It would have been
natural to expect that the $\SL(2,\RR)$ isometries provide the
global part of the associated Virasoro algebra.   The surprise is
that this is not the case.  Specifically, the $\SL(2,\RR)$
isometries are associated to reparameterizations of the non-compact
coordinate $v$ on the boundary, while the physical states only carry
momentum along the compact null direction $u$ on which only the
$\U(1)$ part of the isometry group acts.  Thus, AdS/CFT is telling
us that physical states cannot be charged under the $\SL(2,\RR)$
isometry group associated to the \ads{2} base in
(\ref{NH-BTZ-Sl2xU1}).

Why would a consistent quantum theory of gravity around the
near-horizon BTZ background (\ref{NH-BTZ-Sl2xU1}) require the
absence of excitations in the \ads{2} base of this geometry?
Perhaps because any such fluctuations would cause the space to
``fragment" leading to the appearance of multiple boundaries to the
spacetime \cite{AdS2-fragment}.  In the next section we will
compactify (\ref{NH-BTZ-Sl2xU1}) and examine its stability to
excitations in the \ads{2} base.  Below we will simply accept the
lesson from the analysis of the dual DLCQ field theory and implement
boundary conditions for the spacetime that preserve only the
predicted spectrum.

\noindent {\bf Boundary conditions: }To this end, we will follow the
asymptotic symmetry group analysis of Brown and Henneaux
\cite{Brown-Henneaux} by identifying the boundary conditions for
``allowed'' metric fluctuations close to the spacetime boundary.
First recall the  Brown-Henneaux boundary conditions for \ads{3}. In
the $\hat u,\hat v,r$ coordinates \cite{Carlip}, where the
background  AdS$_3$ metric takes the form $ ds^2=\ell^2(
\frac{dr^2}{r^2}-2 r^2 d{\hat u} d{\hat v}) $
these boundary conditions at large $r$ are \cite{Brown-Henneaux}%
\be\label{Brown-Henneaux-b.c.}%
 \delta g_{\hat u\hat u}\sim \delta g_{\hat v\hat v}\sim
\delta g_{\hat u\hat v}\sim {\cal O}(1), \qquad \delta g_{rr}\sim
{\cal O}(\frac{1}{r^4}),\quad \delta g_{r\hat u}\sim \delta g_{r\hat
v}\sim {\cal
O}(\frac{1}{r^3})\ .%
\ee%
Order one fluctuations in $\delta g_{\hat u\hat u}$, $\delta
g_{\hat v\hat v}$ correspond to normalizable modes in the dual
$2d$ CFT and these may be chosen arbitrarily.  For example,
writing   a generic BTZ black hole in the $\hat u,\hat v$
coordinates, the constant parts of $g_{\hat u\hat u}$ and $g_{\hat
v\hat v}$ determine the ADM mass and angular momentum of the black
hole \eqref{BTZ-uv-ccordinates}. Thus, order ${\cal O}(1)$
fluctuations in $\delta g_{\hat u\hat u}$, $\delta g_{\hat v\hat
v}$ correspond to changing the mass and angular momentum in the
dual $2d$ CFT. A general deformation of $\delta g_{\hat u\hat u}$,
$\delta g_{\hat v \hat v}$ would be non-extremal and would thus
excite both chiral sectors of the dual CFT.  By contrast, we want
to restrict to extremal excitations.  Recalling the form of BTZ
metric \eqref{BTZ-uv-ccordinates}, one may easily observe that
imposing the extremality condition $L_0=c/24$ requires a more
stringent boundary condition on the variations in $g_{\hat v \hat
v}$. The arguments of Sec. \ref{BTZ-dual-CFT} and
\ref{DLCQ-section} for taking the DLCQ limit and in particular
\eqref{Ppm-limit} then suggest that we should replace the boundary
condition on $g_{\hat v \hat
v}$ by%
\be\label{new-gvv}%
\delta g_{\hat{v}\hat{v}}\sim {\cal O}(\frac{1}{r^2}) \,. %
\ee%
 The remainder of the Brown-Henneaux boundary conditions in
\eqref{Brown-Henneaux-b.c.} can be kept intact.  Further analysis
shows that these are forming a set of consistent boundary
conditions.  In fact this set is equivalent to choosing a subset of
\eqref{Brown-Henneaux-b.c.} that preserve the null nature of the
non-compact coordinate $v$ (up to transformations which are trivial
at large $r$).

\noindent{\bf Asymptotic Symmetry Group: } The asymptotic symmetry
group (ASG) of a spacetime is the set of symmetry transformations
(diffeomorphisms) which preserve the boundary conditions modulo the
set of diffeomorphisms the generators of which vanish (reduce to a
boundary integral) after implementation of the boundary conditions.
Equipped with the above boundary conditions we can
compute the ASG for the case of the near horizon extremal BTZ or the
self-dual orbifold of \ads{3}. We seek diffeomorphisms (vector
fields $\zeta$) whose action on the metric (Lie derivative ${\cal
L}_\zeta g$) generates metric fluctuations compatible with the above
boundary conditions. More mathematically, if $g_{\alpha\beta} =
g^0_{\alpha\beta} + \delta g_{\alpha\beta}$, where
$g^0_{\alpha\beta}$ stands for the asymptotic metric, then one is
looking for vector fields $\zeta$ satisfying
\begin{equation}
  \left({\cal L}_\zeta g\right)_{\alpha\beta} \sim \delta g_{\alpha\beta}\,,
\end{equation}
where the symbol $\sim$ stands for same order of magnitude in the large $r$ expansion sense.

Since our boundary conditions are closely related but more
restrictive than those of Brown-Henneaux \cite{Brown-Henneaux}, we
can use their explicit analysis of the generators of the
asymptotic symmetry group and simply impose the additional
constraint on $\delta g_{vv}$ \eqref{new-gvv} on them. The allowed
diffeomorphisms are%
\begin{subequations}\label{BH-diffeos}%
\begin{align}
\zeta^u & =    2 f(u) +\frac{1}{2r^2}g''(v)
+ {\cal O}(r^{-4})  \\
\zeta^{v} & =  2 g(v) + \frac{1}{2r^2}f''(u)+ {\cal O}(r^{-4}),\\
\zeta^r & =  -r\left(f'(u)+ g'(v)\right) + {\cal O}(r^{-1}) %
\end{align}
\end{subequations}%
\be%
 g'''(v)=0 \quad \Longrightarrow \quad g=A+B\,v + C \, v^2
\label{B-condition}\ .
\ee%
Here, the connection to the Brown-Henneaux diffeomorphisms is made
explicit: the diffeomorphisms generated by $\zeta=\zeta^\alpha
\partial_\alpha$ of \eqref{BH-diffeos} are exactly those of
Brown-Henneaux \cite{Brown-Henneaux} and the constraint $\delta
g_{vv}={\cal O}(\frac{1}{r^2})$ is implemented by
\eqref{B-condition}.  One set of allowed diffeomorphisms is
specified by a periodic function $f(u)=f(u+2\pi)$. The analysis of
generators of these diffeomorphisms follows directly from those of
Brown and Henneaux and they lead to a \emph{chiral Virasoro algebra}
at central charge $c= 3\ell/2G_3$ \eqref{B.H-central-charge}. \footnote{Our analysis here suggests that the Left and Right CFT's introduced in \cite{Marolf} may be identical. This point deserves further investigation.}
 The remaining three parameter family of diffeomorphisms in
(\ref{B-condition}) describes the $\SL(2,\RR)$ isometries of the
self-dual orbifold.

The isometries of the original extremal black holes were just a
$\U(1) \times \U(1)$.  In that case a Brown-Henneaux analysis with
the extremal constraint would have also yielded \eqref{BH-diffeos}
with the constraint $g''' = 0$.  However, in the original geometry
$g$ has to be a periodic function which restricts the solutions to
the constraint to $g = A$ only.   The process of taking the near
horizon limit led to an identification in $u$ alone, and thus, $g$
need not be periodic, allowing the three parameter solution above.
The isometry generators that appear in this way, are not simply
related to the $SL(2,\RR)$ generated   by $L_0,L_{\pm 1}$
\eqref{ads3virasoro}.

\section{$AdS_2$ quantum gravity and dual chiral CFTs}\label{2dEMD-section}

Consider the two-dimensional Einstein-Maxwell-Dilaton theory with a
negative cosmological constant:
\be%
S=\frac{\ell}{8G_3}\int d^2x
\sqrt{-g}\left[e^{\psi}(R+\frac{2}{\ell^2})-\frac{\ell^2}{4}e^{3\psi}
F_{\mu\nu}F^{\mu\nu}\right]%
\Label{2d-action}
\ee%
where $F_{\mu\nu}$ is the $U(1)$ field strength. This action has an \ads{2}
solution with curvature $R=-\frac{8}{\ell^2}$, constant
$\psi$ and constant electric flux :%
\be%
\begin{split}%
ds^2=-\frac{\ell^2}{r^2}(-dt^2+dr^2),\qquad
 F_{tr}=\frac{2Q}{r^2},
\qquad e^{-\psi}=Q\ .
\end{split}%
\Label{AdS2-soln}
\ee%
This action may be obtained from the dimensional reduction of the
$3d$ Einstein-Hilbert action with $3d$ Newton constant $G_3$ and
cosmological constant $-1/\ell^2$ via restriction to the massless
sector of the Kaluza-Klein tower. \footnote{An analysis of
Schwinger pair creation of charged particles in $AdS_2$ in the
presence of a constant electric field was performed in
\cite{boris-jan}. A bound between the mass of particle excitations
and the background electric field was derived to ensure the
stability of these backgrounds. This bound is satisfied in
supersymmetric $AdS_2\times S^2$ spacetimes and is also saturated
for the two dimensional vacuum solution discussed here.} Likewise
the reduction of the near-horizon BTZ geometry
\eqref{NH-BTZ-Sl2xU1} to two dimensions is precisely
(\ref{AdS2-soln}).  The radius of the extremal BTZ horizon becomes
$\ell Q$.   The action (\ref{2d-action}) has another two parameter
family of  solutions in which $\psi$ is not a constant
\cite{AdS2QGr} -- these lift to generic BTZ black holes.

Because of this connection between two and three dimensions, we
expect that quantum gravity around the background \eqref{AdS2-soln}
is  dual to a subsector of the DLCQ chiral CFT that is developed in
Sec.~\ref{BTZ-dual-CFT} and \ref{DLCQ-section}, and is only fully
consistent when embedded in string theory with all the resulting
additional degrees of freedom.  The electric field strength $Q$ is
related to the DLCQ compactification scale $R_-$ in
(\ref{tpmy-periodicity}) while the central charge is related to the
$2d$ Newton constant: $c=3\ell/(2G_3)=3/(4\pi G_2)$.

Quantum gravity in the \ads{2} background (\ref{AdS2-soln}) was
explored in \cite{Hartman} from the perspective of the spacetime
conformal field theory, and in \cite{Finn-Larsen} from the
perspective of the boundary stress tensor.   Both of the papers
consider spectra including states charged under the $\SL(2,\RR)$
isometry group of \ads{2}, and analyze a Virasoro algebra which
includes this $\SL(2,\RR)$.   However, as shown in previous
sections, a consistent quantum theory of gravity in this background
should not have any states charged under the isometry group.  The
reason for this is that excitations supported in \ads{2} back-react
strongly and can modify the asymptotic structure of the spacetime
\cite{AdS2-fragment}.

To see this, let us write the two dimensional metric in a gauge in which the metric is conformally flat%
\be
ds^2=e^{2\phi(\sigma^+,\sigma^-)} \, d\sigma^+d\sigma^-\ \ , \ \ 0\leq \sigma^{\pm} \leq \pi \ \ ,
\ee
and consider the variation of the action \eqref{2d-action} with
respect to the $2d$ metric. We find
\be \label{aux22}%
\nabla_+ \nabla_+ e^{\psi}  =  8\pi G_2 T_{++} %
\ee %
and similarly for the $--$ component.  If we regard
(\ref{2d-action}) as arising from compactification of a three
dimensional theory, besides the contributions from $\psi$ and the
gauge field, we can also include all contributions of massive
Kaluza-Klein modes in $T_{++}$.   We may now follow the discussion
in section 2.2 of \cite{AdS2-fragment} (see Eqs.~(2.16) and (2.17)
there). Integrating (\ref{aux22}) against
$e^{-2\phi}d\sigma^+$, we obtain%
\be%
 e^{-2\phi} \partial_+
e^{\psi} |_{\sigma^+=0} - e^{-2\phi}
\partial_+ e^{\psi} |_{\sigma^+=\pi} = -8\pi G_2 \int d\sigma^+ e^{-2\phi}
T_{++} %
\ee%
and similarly for $T_{--}$. Assuming a null energy condition
($T_{++} \geq 0$), any state with non-vanishing $T_{++}$ requires at
least one of the two terms on the left hand side of this equation to
be non-zero.    Since $e^{-2\phi}$ vanishes quadratically near the
boundary of AdS$_2$, this implies that $e^\psi$ must diverge at one
of the \ads{2} boundaries.  This is inconsistent with the constant
value $e^{-\psi} = Q$ in (\ref{AdS2-soln}), which is related from
the three dimensional point of view to the compactification radius.
This shows that preserving the boundary conditions  requires $T_{++}
= 0$ and a similar argument requires $T_{--}= 0 $.  Thus
perturbations cannot have any dependence on $\sigma^+,\sigma^-$, as
their back-reaction would destroy the boundary of the geometry. The
background geometry \eqref{AdS2-soln} has an $\SL(2,\mathbb R)$
isometry, and if perturbations do not depend on $\sigma^+,\sigma^-$,
then the perturbation cannot break the $\SL(2,\mathbb R)$ symmetry
either. In other words, all degrees of freedom transform trivially
under $\SL(2,\mathbb R)$, in agreement with the analysis in previous
section.\footnote{The fact that \ads{2} ``fragments'' in this way
has led to the suggestion \cite{AdS2-fragment} that the dual of a
one-dimensional conformal field theory should involve a sum over
tree-like geometries with many different asymptotic \ads{2}
boundaries. While some partial progress has been made in developing
this picture \cite{michelsonstrominger,ooguridijkgraafvafa}, it is
still unclear whether this is the right way to think about \ads{2},
or whether it eventually will lead to connection with the fuzzball
proposal, and we will not further pursue this possibility in this
paper.}

This argument used the fact that \ads{2} has two disconnected
boundaries.  In Sec.~\ref{BTZ-dual-CFT}  the analysis of the CFT
dual was carried out in coordinates that only intersected a single
boundary, but it was shown in \cite{Vijay-Asad-Joan} that,
globally, the self-dual orbifold geometry has two boundaries, each
of which is a null cylinder carrying a DLCQ of a CFT. To see this,
transform the
coordinates in (\ref{NH-BTZ-Sl2xU1}) as %
\be%
y =  \cos\tau \, \cosh z
+ \sinh z ~~~~;~~~~ v = {\sin \tau \, \cosh z \over \cos \tau \,
\cosh z + \sinh z} \, , \Label{globaltransf} %
\ee%
so that the self-dual orbifold metric becomes%
\be%
 ds^2 = {\ell^2
\over 4} \left( -\cosh^2 z \, d\tau^2 + dz^2 \right) + \ell^2
\left(du + A' \right)^2 \,%
 \ee%
where $A'$ is a gauge field with constant field strength in global
\ads{2}.  This is the global self-dual orbifold of
\cite{Vijay-Asad-Joan}.  The entire range of $v$ is covered by a
finite range of global time $\tau$.   Thus each patch of the form
(\ref{NH-BTZ-Sl2xU1}) intersects one boundary of the global
spacetime at either $z = \pm \infty$.

In view of this, both the near-horizon limit of extremal BTZ
(\ref{NHlimit-BTZmet1}) and the $3d$ uplift of (\ref{AdS2-soln}) can
be regarded {\it globally} as dual to {\it two} DLCQ CFTs, each
giving rise to one chiral theory (see \cite{Vijay-Asad-Joan,tak} for
discussion).    From this perspective  we can presumably view the
description of the self-dual orbifold as a thermal state in a single
CFT as emerging from tracing over the Hilbert space living in one of
the boundaries.   This is in analogy with the usual treatment of the
eternal BTZ black hole as either an entangled state in two CFTs
defined on the two boundaries of the geodesically complete
spacetime, or as a thermal state in a single CFT
\cite{Maldacena:2001kr}.  The statistical degeneracy of the thermal
state in the chiral CFT dual to the spacetime (\ref{NH-BTZ-Sl2xU1})
then measures the area of the familiar Poincare horizon of this
coordinate patch (see \cite{Spradlin:1999bn} for a similar
perspective).    One difference between BTZ and the self-dual
orbifold is that while the BTZ boundaries are causally disconnected,
a light ray can travel between the two boundaries of the global
self-dual orbifold \cite{Vijay-Asad-Joan}.  The possible
interactions that this seeds between the two boundaries have not
been studied.

In this global context there is another piece of evidence that the
\ads{2} base of the self-dual orbifold cannot be consistently
excited.   It was shown in \cite{Finn-Larsen} that the most
general solution of the dimensional reduction of $3d$ gravity with
a negative cosmological constant in a particular gauge can be put
in the form
\be%
g_{\mu\nu} dx^{\mu} dx^{\nu} = d\eta^2 - \frac{1}{4} (h_0(t)
e^{2\eta/L} + h_1(t) e^{-2\eta/L})^2 dt^2. %
 \Label{eq210}
\ee%
 At the boundary $\eta\rightarrow \infty$,
the boundary metric is determined by $h_0(t)$. One can choose a
coordinate $t$ such that $h_0=1$. The subleading behavior is
determined by $h_1(t)$.  The diffeomorphisms that preserve this
gauge and leave $h_0$ unchanged were determined by
\cite{Finn-Larsen}. However it turns out that while these are
normalizable deformations of the boundary at $\eta\rightarrow
+\infty$, they are not normalizable deformations at the other
boundary $\eta\rightarrow -\infty$ -- i.e. they change $h_1$.   In
fact, there are no deformations at all which both preserve the gauge
and are normalizable at both boundaries, except the isometries.
This again suggests that it is not possible to deform \ads{2}
without disrupting the spacetime boundary.

\section{Extremal Kerr black hole and its dual chiral CFT}

The extremal $4d$ Kerr black hole is given by%
\be \label{Ext-Kerr}%
ds^2=-\frac{\Delta}{R^2}\left(d\hat t-a\sin^2\theta
d\hat\phi\right)^2+\frac{\sin^2\theta}{R^2}\left((\hat
r^2+a^2)d\hat\phi-ad\hat t\right)^2+\frac{R^2}{\Delta}d\hat
r^2+R^2d\theta^2\ , %
\ee%
where%
\be%
R^2=\hat r^2+a^2\cos^2\theta,\qquad \Delta=(\hat r-a)^2\ . %
\ee%
Its ADM mass and angular momentum are function of the horizon size $a$%
\be\label{ADM-MJ-Kerr}%
M=a,\qquad J=\frac{a^2}{G_4}\ .%
\ee%
In the quantum theory, $J$ is quantized (to half integers) in units
of $\hbar$. This black hole has zero Hawking
temperature and its Bekenstein-Hawking entropy is%
\be\label{BH-entropy-Ext-Kerr}%
S_{BH}=\frac{2\pi M^2}{\hbar G_4}=\frac{2\pi}{\hbar}\ J\ . %
\ee%

In the near horizon $\epsilon\to 0$ limit%
\be%
\hat r=a+\epsilon\ r,\quad \hat t=\frac{2at}{\epsilon},\quad
\hat\phi=\phi+\frac{t}{\epsilon}, %
\ee%
while keeping the un-hatted parameters and coordinates fixed, we
obtain the near horizon extremal Kerr (NHEK) geometry
\cite{Bardeen-Horowitz, Kerr/CFT}%
\begin{equation}\label{NHEK}%
  ds^2 = 2 G_4J\,\Omega(\theta)^2\left[-r^2 dt^2 + \frac{dr^2}{r^2}+ d\theta^2 +
  \Lambda(\theta)^2\,\left(d\varphi + r dt\right)^2\right]\,,
\end{equation}
where $\varphi\in [0,2\pi],\ 0\leq \theta\leq \pi$ and%
\be%
\Omega(\theta)^2 = \frac{1+\cos^2\theta}{2},\qquad
\Lambda(\theta)=\frac{2\sin\theta}{1+\cos^2\theta}. %
\ee%
This metric at a given $\theta$ has the form of a warped circle
fibration over \ads{2} in which the fiber radius depends on the
angle $\theta$. If $\Lambda$ and $\Omega$ were constants this would
be precisely the self-dual orbifold of (\ref{NH-BTZ-Sl2xU1}) times a
circle. Indeed, as emphasized in \cite{Kerr/CFT}, constant $\theta$
slices look like squashed self-dual orbifolds.   The coordinates in  (\ref{NHEK}) cover only part of the spacetime, with a boundary at $r \to \infty$ -- globally, like the self-dual orbifold, there are two boundaries.   One sees similar
squashed geometries with \ads{2} and \ads{3} factors in decoupling
limits of near-extremal black holes in anti-de Sitter space
\cite{Balasubramanian:2007bs,Fareghbal:2008ar,Fareghbal:2008eh}.
(Also see \cite{Ext/CFT-1,Ext/CFT-2, Kerr/CFT-Strings}.)

The  Kerr black hole is invariant under time and angular $\hat \phi$
translations. This isometry group is enhanced to
$\SL(2,\mathbb{R})\times \U(1)$ in the near horizon, just as in the
self-dual orbifold.  The $\U(1)$ is generated by $\partial_\varphi$,
whereas the $\SL(2,\mathbb{R})$ acts both on the AdS$_2$ subspace
and along the fiber to preserve the form of $d\varphi+ r dt$
\cite{Kerr/CFT}.

In \cite{Kerr/CFT}, the asymptotic symmetry group preserving certain  boundary conditions for the fluctuations of  the
NHEK was calculated. The corresponding diffeomorphisms they found were of the form%
\be%
\zeta_\lambda=\lambda(\varphi)\partial_\varphi-r
\lambda(\varphi)' \partial_r\,.%
\ee%
These generate a chiral Virasoro algebra.  In \cite{Kerr/CFT} it was
proposed that this Virasoro algebra should be understood as the
symmetry group of a chiral $2d$ CFT dual to quantum gravity around
the near horizon Kerr geometry.
The central charge of this chiral CFT was computed to be%
\be\label{c-Kerr}%
c_{Ext.\ Kerr}=12 J\ .%
\ee%
The NHEK is then associated with a thermal state of the chiral $2d$
CFT at temperature $T_{NHEK}=1/2\pi$. Upon applying the Cardy
formula for the entropy of $2d$ CFTs, the Bekenstein-Hawking entropy
of the extremal Kerr black hole \eqref{BH-entropy-Ext-Kerr} is
reproduced. The consistency of the boundary conditions proposed in
\cite{Kerr/CFT} required the vanishing of  the charge of the
$U(1)_\tau\in SL(2,\mathbb{R})$, i.e. \be E_R = 0 \ee in the
notation used in \cite{Kerr/CFT}, for all physical states. Thus,
like for the self-dual orbifold, there are no physical excitations
of the \ads{2} factor in the geometry.  The $E_R = 0$ condition acts
like the restriction to extremality in the BTZ black hole that we
studied in Sec.~2.

The analogies between the Kerr-CFT construction \cite{Kerr/CFT} and
the analysis of the self-dual orbifold in previous sections suggests
that chiral CFT of \cite{Kerr/CFT} is the DLCQ of an ordinary two
dimensional conformal field theory.    Ideally, we would like to
find a consistent Kaluza-Klein reduction of gravity in the NHEK
geometry to the three-dimensional self-dual orbifold.     As a first
step, we make a connection between the NHEK geometry and $3d$
gravity with a negative cosmological constant. For the NHEK geometry we consider then the four dimensional
metric reduction ansatz:%
\begin{equation}\label{reduction-4to3}
  ds^2 = L^2\,\Omega^2\left[-\partial_\sigma\beta (t,\,\sigma)\,\left(-dt^2 + d\sigma^2\right) + d\theta^2 + \Lambda^2\,\left(d\varphi + \beta (t,\,\sigma) dt\right)^2\right]\,,
\end{equation}
where $\Omega^2 = (1+\cos^2\theta)/2$ and
$\Lambda=2\sin\theta/(1+\cos^2\theta)$.   The equation of motion derived for $\beta$ using this ansatz and the four dimensional Einstein equation {\it without} a cosmological constant is identical to the equation of motion obtained from the three-dimensional ansatz
\be%
\label{aux12} ds^2= \frac{\ell^2}{4} \left[-\partial_\sigma\beta
(t,\,\sigma)\,\left(-dt^2 +
d\sigma^2\right)+\left(d\varphi + \beta (t,\,\sigma) dt \right)^2\right]\ . %
\ee%
and Einstein's equation {\it with} a cosmological constant
\be\label{3d-eom}%
R_{3}{}_{\mu\nu}+\frac{2}{\ell^2} g_{3}{}_{\mu\nu}=0 . %
\ee%
Here $R_3$ is the Ricci tensor computed for the $3d$ metric.
Although this obviously does not show that there should exist a
Kaluza-Klein reduction from four to three dimensions which reduces
the NHEK geometry to the self-dual orbifold of \ads{3}, it does show
that the two theories share some dynamics.

We can also derive the central charge derived in \cite{Kerr/CFT}
from the $4d$ NHEK geometry, by matching parameters with the
three-dimensional reduction ansatz.
To do this, note first that the above $3d$ equation of motion can be obtained from the Lagrangian%
\be%
{\cal L}_3=\sqrt{-\det g_3}(R_3+\frac{2}{\ell^2})\,, %
\ee%
which describes $3d$ gravity in the presence of a negative
cosmological constant.
The $3d$ Newton constant is then computed by integrating over the compact direction $\theta$ in our reduction ansatz%
\be\label{GN-3d}%
\frac{1}{G_3}=\frac{2L^2 \int^\pi_0 d\theta\ \Omega^2\Lambda}{G_4\
\ell}=
\frac{4L^2}{G_4\ \ell}\,.%
\ee%
Thus the $3d$ action is%
\be\label{reduced-3d-action}%
S_3=\frac{1}{16\pi G_3}\ \int d^3x \ {\cal L}_3\ , %
\ee%
Note that its vacuum solution is an AdS${}_3$ with radius
$R_{AdS}=\ell$. Since $L^2=2G_4 J$, using the
Brown-Henneaux formula for the central charge, we have%
\be%
c=\frac{3 R_{AdS}}{2 G_3}=12 J\,. %
\ee%
This matches  \eqref{c-Kerr}.  We earlier showed that the \ads{3}
central charge also matches the central charge of the chiral CFT that is
dual to self-dual orbifold.

This suggests  the proposed chiral $2d$ CFT dual to extremal Kerr
\cite{Kerr/CFT} is the DLCQ of a $2d$ CFT with the following
identifications: (a)  The DLCQ compactification radius $R_-$ is an
arbitrary physical scale and has been set equal to one in the
Kerr/CFT analysis \cite{Kerr/CFT}, (b) The $E_R=0$ condition in
\cite{Kerr/CFT} is  mapped to $L_0=c/24$ DLCQ condition, (c) The
extremal Kerr ADM angular momentum $J$ is equal to the light-cone
momentum $P^+$ of the DLCQ description.

One should note that identifying the chiral $2d$ CFT duals proposed
for extremal black holes \cite{Kerr/CFT,Hartman:2008pb} as the DLCQ
of a $2d$ CFT also explains why we can use Cardy's formula to count
the number of states. If we only knew that the states had to form
representations of a single Virasoro algebra, we would not be able
to use modular invariance, and unitarity alone does not determine
the asymptotic growth of the number of states.  Still, there are to
our knowledge no general statements about the asymptotic growth of
the number of states of the form $|c/24\rangle_R \otimes |{\rm
anything}\rangle$ in an arbitrary CFT. If the left-movers are Ramond
ground states, and it is a theory with supersymmetry, one can
estimate the number of states of this form using the elliptic genus
and its modular properties \cite{elliptic-genus}, and it would be
interesting to establish similar results for more general CFT's.

While our results have provided some evidence that  DLCQ of a CFT is
dual to the near-horizon extremal Kerr, it would have been more
satisfactory to have a consistent and complete reduction of $4d$
gravity with NHEK boundary conditions \cite{Kerr/CFT} to $3d$
gravity with a cosmological constant.  In a similar setting where
squashed \ads{3} factors appear in a decoupling limit of R-charged
black holes in \ads{4} and \ads{5}, progress towards such a
reduction has been made
\cite{Balasubramanian:2007bs,Fareghbal:2008ar,Fareghbal:2008eh}.

\section{Discussion}

In this paper we have shown that the near-horizon limit of the extremal BTZ
black hole, which leads to the so-called self-dual orbifold
geometry, is dual to the DLCQ of a non-chiral 2d CFT, which is
a {\it chiral 2d CFT} with the same central charge. We
have also provided evidence that various ``chiral CFTs'' that have
appeared in the literature as dual CFTs to extremal black holes
should really be thought of as DLCQ of ordinary two-dimensional
CFTs. This, among other things, justifies the use of Cardy formula
to account for the extremal black hole entropy using this chiral CFT
duals. It would be desirable to develop this picture in more detail.
In particular, it would be interesting to study correlation
functions in the DLCQ theory and the corresponding bulk-boundary
dictionary. Another outstanding problem is to establish more
rigorously that generic extremal black holes, upon taking a
near-horizon limit, are indeed dual (once suitable boundary
conditions are imposed) to the DLCQ of a conformal field theory. If
this is indeed the case, one would expect that the parent $2d$ CFT
of the DLCQ theory might also have a string theoretic realization,
e.g. in the form of a warped \ads{3} solution of string theory. In
other words, one might seek some sort of map from extremal black
hole solutions to \ads{3} solutions. We have seen hints of such a
map in
\cite{Balasubramanian:2007bs,Fareghbal:2008ar,Fareghbal:2008eh}, but
whether it exists in the general case is unclear.

One curiosity about the self-dual orbifold geometry is that it is
dual to thermal state in a DLCQ CFT.   The {\it ground state} of the
DLCQ theory does not appear to have a bona fide geometric dual.\footnote{Specifically, it is {\it not} dual to a very near horizon limit of the $M=0$ BTZ black hole as one can explicitly check.}
This is unlike \ads{3} gravity with a standard cylindrical boundary
where the ground state describes empty AdS and thermal states
describe black holes.   This seems to be a general feature of
gauge-gravity duality for DLCQ field theories \cite{CMT}.

\section*{Acknowledgements}
We would like to thank John McGreevy, Bindusar Sahoo, Gary Horowitz and Matthew
Roberts for useful discussions. J.S. would like to thank the
organizers of the KITP programme``Fundamentals of String Theory" and
the Departmento de F\'{\i}sica de Part\'{\i}culas in Santiago de
Compostela for hospitality during different stages of this project.
The work of J.S. was partially supported by the Engineering and
Physical Sciences Research Council [grant number EP/G007985/1]. This
research was supported in part by the National Science Foundation
under Grant No. NSF PHY05-51164.  VB was partly supported by  DOE
grant DE-FG02-95ER40893 and partly by NSF grant NSF OISE-0443607.
JdB was supported in part by the FOM foundation.


\providecommand{\href}[2]{#2}\begingroup\raggedright

\endgroup

\end{document}